# Mesosphere Study by Wide-Field Twilight Polarization Measurements: First Results beyond the Polar Circle

Oleg S. Ugolnikov[1], Boris V. Kozelov[2]

[1]Space Research Institute, Russian Academy of Sciences, Moscow, Russia
[2]Polar Geophysical Institute, Apatity, Russia

E-mail: ougolnikov@gmail.com, boris.kozelov@gmail.com

**Abstract**
The paper describes the observations and first results of studying the mesosphere temperature and dust based on twilight wide-field polarization analysis using original all-sky camera started in 2015 in Apatity, northern Russia (67.6°N, 33.4°E). It is the first twilight polarization measurements session conducted in the polar region and the first measurements during the winter and early spring season. The general polarization properties of twilight sky and the separation procedure of single scattering are described. The Boltzmann temperature decrease above 70 km and lack of mesosphere dust that is typical for this season are the key results of the study.

## 1. Introduction

Mesosphere of the Earth is one of the coldest regions in the atmosphere and on the entire planet, especially in the polar latitudes during the summer where the temperature can drop down to 120K and below [Höffner and Lübken, 2007; Dyrland et al., 2010]. The inverted annual temperature cycle with a summer minimum and an amplitude of up to 100K is another remarkable feature of the mesosphere. This is due to the fact that the mesosphere is in the thermal non-equilibrium state, being highly sensitive to various chemical and physical factors. This peculiarity strongly impedes the measurements of the mesosphere temperature and can lead to the inconsistencies in the results provided by different methods.

The temperature studies of the mesosphere are of special interest since it accommodates the fastest climate changes during the recent decades, which are probably related to the increased level of greenhouse gases, especially $CO_2$ [Houghton, 1970; Roble and Dickinson, 1989]. The measurements by the Soviet meteorological rocket M-100 during the second half of 20th century [Golitsyn et al, 1996] have shown that the mean annual temperature could have decreased by almost 1K per year, i.e. severalfold faster than any changes observed in the lower atmosphere. The trend values increased in the polar regions. The present methods and trend estimates for different altitudes and latitudes are reviewed in [Beig et al., 2003; Beig, 2006, 2011ab].

At the same time, the mesosphere is also the least studied atmospheric layer. Most information for this region is obtained by remote sensing from ground or from space. Satellites provide global data covering a wide altitude range. At present, two space instruments, SABER onboard the TIMED satellite [Russell et al., 1999] and MLS onboard the EOS Aura satellite [Schwartz et al., 2008] are measuring the temperature in the mesosphere based on $CO_2$ and $O_2$ spectroscopy, respectively.

The ground-based techniques provide local data for a certain altitude range. The OH spectroscopy in Meinel bands gives the temperature values around to the summer mesopause (82-89 km), the layer of maximal OH concentration [Baker and Stair, 1988; Pertsev and Perminov, 2008]. Meteor radar data [Dyrland et al., 2010] are used for finding the temperature near 90 km; the molecular oxygen airglow measurements are related to the altitude about 95 km [Reisin and Scheer, 2002]. Sodium [She et al., 1982] and potassium [Höffner and Lübken, 2007] lidars retrieve the temperature profiles inside the metal layers, from 80 to 110 km. The profiles can also be obtained by the Rayleigh lidars [Hauchecorne and Chanin, 1980]. A "falling sphere" tracing [Lübken and Müllemann, 2003] as another effective albeit quite expensive technique.

There are rather limited stations for local mesospheric measurements, especially in polar latitudes where temperature variations and probable trends are highest. The potassium lidar [Höffner and Lübken, 2007] and meteor radar [Dyrland et al., 2010] observations at Svalbard, Norway (78°N) are worthy of note.

The main objective of the present work is to install the instrument for mesospheric temperature measurements in the polar region of Russia, which has still not been covered by such stations. For this purpose, we chose the twilight sounding technique, which is least expensive. This method is efficient if the intensity and polarization of the twilight background are measured simultaneously at many sky points. This task was successfully achieved in the mid-latitude Russia [Ugolnikov and Maslov, 2013ab, 2014, 2015]. The temperature in the upper mesosphere (70-85 km) was measured with high accuracy and in agreement with the satellite data. The polarization study enables the multiple scattering background to be separated and the dust presence in the mesosphere to be detected. This was done for the Perseids activity epoch [Ugolnikov and Maslov, 2014].

## 2. Observations

The polarization twilight sky measurements were started in February 2015 at the Polar Geophysical Institute, Apatity, Russia (67.6°N, 33.4°E). The main optical instrument was the OSSH-08-GAO lens (Kiev, Ukraine), which creates the all-sky image with a field of view of about 180°. In the optical scheme of the lens, the rays are collimated into a small axial angle, allowing the use of a Sony circular polarizer filter (CPL). The original mechanics can rotate the filter around the optical axis of the lens with a speed up to 30 degrees per second. The detailed descriptions of the hardware and the data storage procedures will be presented in a separate paper.

The color filters created a wide spectral band with an effective wavelength of 530 nm and FWHM about 70 nm. This band is close to that used in the similar measurements in central Russia (540 nm, [Ugolnikov and Maslov, 2013ab]). This spectral region is most suitable for the Rayleigh scattering analysis since it is free from gaseous absorption (except for the wide ozone Chappuis bands) and is moderately affected by the tropospheric multiple scattering (increasing bluewards) and aerosol (increasing redwards). The same wavelength was used for the Rayleigh scattering analysis in the WINDII experiment onboard the UARS satellite [Shepherd et al., 2001].

The emission field was fixed by the Prosilica GC-780 CCD camera. The wide range of the exposure times (from 100μs to 10s) allowed us to cover the whole twilight period and to correct the CCD saturation effects. Since the observations were conducted in the light-polluted town sky, the data at the points of the sky area with zenith distances of up to 65° were taken into account. The same limitation was applied in [Ugolnikov and Maslov, 2013ab] in central Russia.

In this paper we present the very first results of the study. The initial observation session covered February and March 2015, followed by no clear skies period in April, and a light polar twilight after it. However, the obtained data are sufficient to demonstrate the efficiency and possibility of twilight sounding for the polar mesosphere study. This period is transitional between winter and summer, when the mesosphere temperature is expected to decrease with time.

## 3. Twilight sky polarization properties

The dependency of the sky polarization on the solar zenith angle during the twilight reveals the morphological changes in the background composition and can serve as the basis for subdividing the twilight period into the separate stages [Ugolnikov and Maslov, 2007]. Figure 1 shows the dependences of the normalized second Stokes component on a solar zenith angle for a number of solar vertical points in the morning twilight of February, 14. This value is negative if polarization is directed towards the Sun and positive in the case of perpendicular direction. These points are characterized by the zenith distance $\zeta$, positive in the dawn area and negative in the opposite part of the vertical.

The curves are quite similar to those observed in the mid-latitudes [Ugolnikov and Maslov, 2013ab]. The dark stage with a solar zenith angle $z_0>97°$ is the most interesting period of the twilight, when the mesosphere can be studied. During this interval, the local troposphere and stratosphere are fully immersed into the Earth's shadow and are not illuminated by the direct solar emission, but the mesosphere still is. However, the lower atmospheric layers affect the sky background by means of multiple light scattering. The contribution of multiple scattering is significant during the whole twilight stage and it is the cause of "reverse polarization" far from zenith as it is seen in Figure 1. Multiple scattering becomes predominant during the dark period of the twilight, especially in the dawn-opposite area where the Earth's shadow ascends very high above the ground. The dawn segment still contains the higher-polarized single scattering, which introduces asymmetry into the polarization along the solar vertical during this stage of twilight, as can be easily seen in Figure 1.

During the darker stage, at a solar zenith angle of about 100°, single scattering totally disappears in the whole observable part of the sky. The polarization becomes symmetric along the solar vertical with a maximum at the zenith, which can be treated as a property of multiple scattering [Ugolnikov, 1999; Ugolnikov and Maslov, 2002]. In the areas with weak light pollution and night sky background, the polarization remains constant up to $z_0$ of about 102°; but in the urban conditions the polarization starts to vanish this time.

The method for separating the single and multiple scattering suggested in [Ugolnikov and Maslov, 2013b] constructs the linear relationship between the Stokes vectors of multiple scattering in the symmetric sky points during the darkest twilight stage ($z_0>100°$), which is then used for finding the single scattering field in the dusk/dawn area for $97°<z_0<100°$.

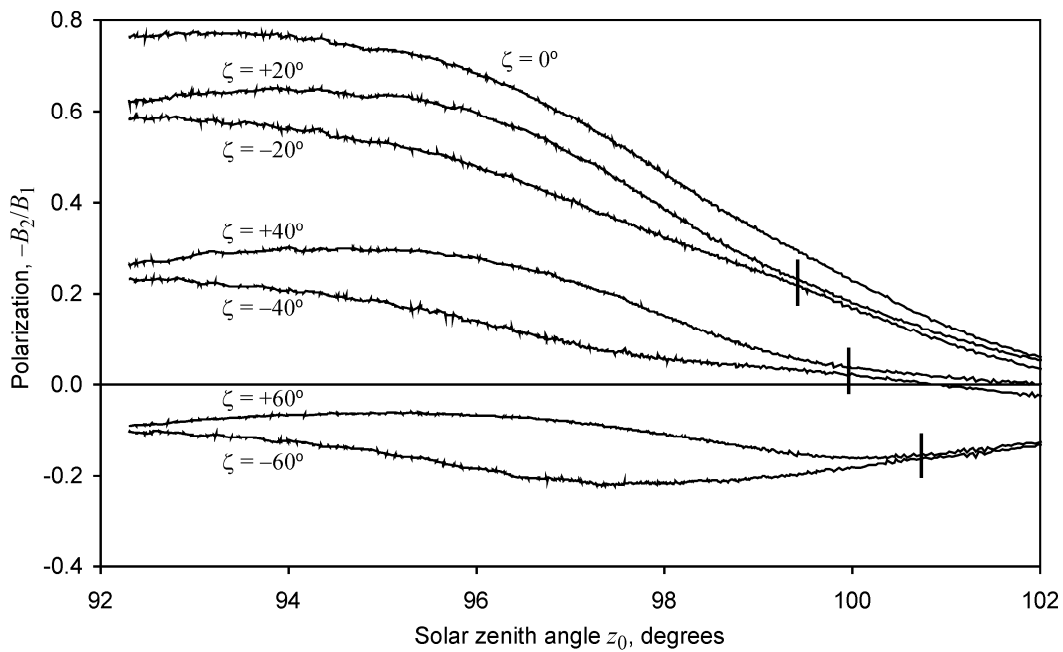

*Figure 1. Normalized second Stokes component in the solar vertical points during the morning twilight of February, 14, 2015. Short vertical lines show the moments of single scattering disappearance in the dawn segment.*

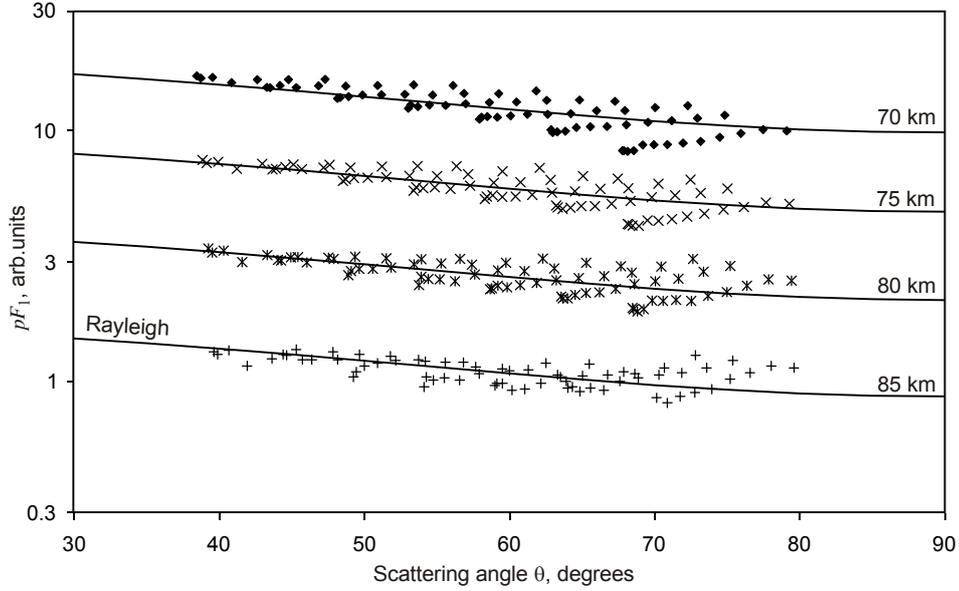

*Figure 2. Single scattering functions for different effective baseline altitudes, the morning twilight of February, 14, 2015.*

The mathematical procedure is described in the cited paper. The keystone of this procedure is the value of $z_{0S}$ corresponding to the situation when the single scattering completely disappears (as a function of the sky point coordinate $\zeta$, equal to the zenith distance in the solar vertical). This value depends on observation site and wavelength and it should be determined from the polarization dependences like those shown in Figure 1. The analysis shows that the empirical formula for these observations is similar to that in [Ugolnikov and Maslov, 2013b] with a small correction:

$$z_{0S} = 99.0 + 0.02 \cdot \zeta + 0.0000025 \cdot \zeta^3 \qquad (1).$$

Here, all values are expressed in degrees. The last term is significant far from zenith and it eventually shifts the mesospheric temperatures by about 5K These $z_{0S}$ values for $\zeta$ of 20°, 40° and 60° are shown on the curves in the Figure 1.

## 4. Temperature analysis

The single-scattering brightness of a sky point measured during the twilight is an integral value contributed by the different atmospheric layers. We may assume it to be proportional to the concentration integral above a certain altitude or corresponding pressure at this altitude [Ugolnikov and Maslov, 2013b]. However, due to the strong extinction of tangent emission above the Earth's surface, this altitude is not equal to that of the geometrical shadow. The difference can be found from the lower atmosphere model for the observation point; however, its variation by the location and season is rather small. Ugolnikov and Maslov [2013] calculated that the effective layer baseline altitude for summer in central Russia at 540 nm during the dark twilight corresponds to the tangent solar ray with a perigee height equal to 14.2 km. For wintertime at 530 nm, it is slightly higher, 16.0 km, and the refraction of these rays should be taken into account.

Figure 2 shows the single scattering functions (in arbitrary units) for the effective layer baseline altitudes 70, 75, 80, and 85 km obtained by the procedure described above for the same morning twilight of February, 14, 2015. These functions are close to the Rayleigh ones shown by the solid lines (this will be also supported below by the polarization analysis). Assuming that these values of $pF_1$ are proportional to the pressure at the same altitude $p$ and Rayleigh scattering function per molecule $F_1$, we build the pressure and then temperature profiles in the upper mesosphere (see [Ugolnikov and Maslov, 2013b] for details).

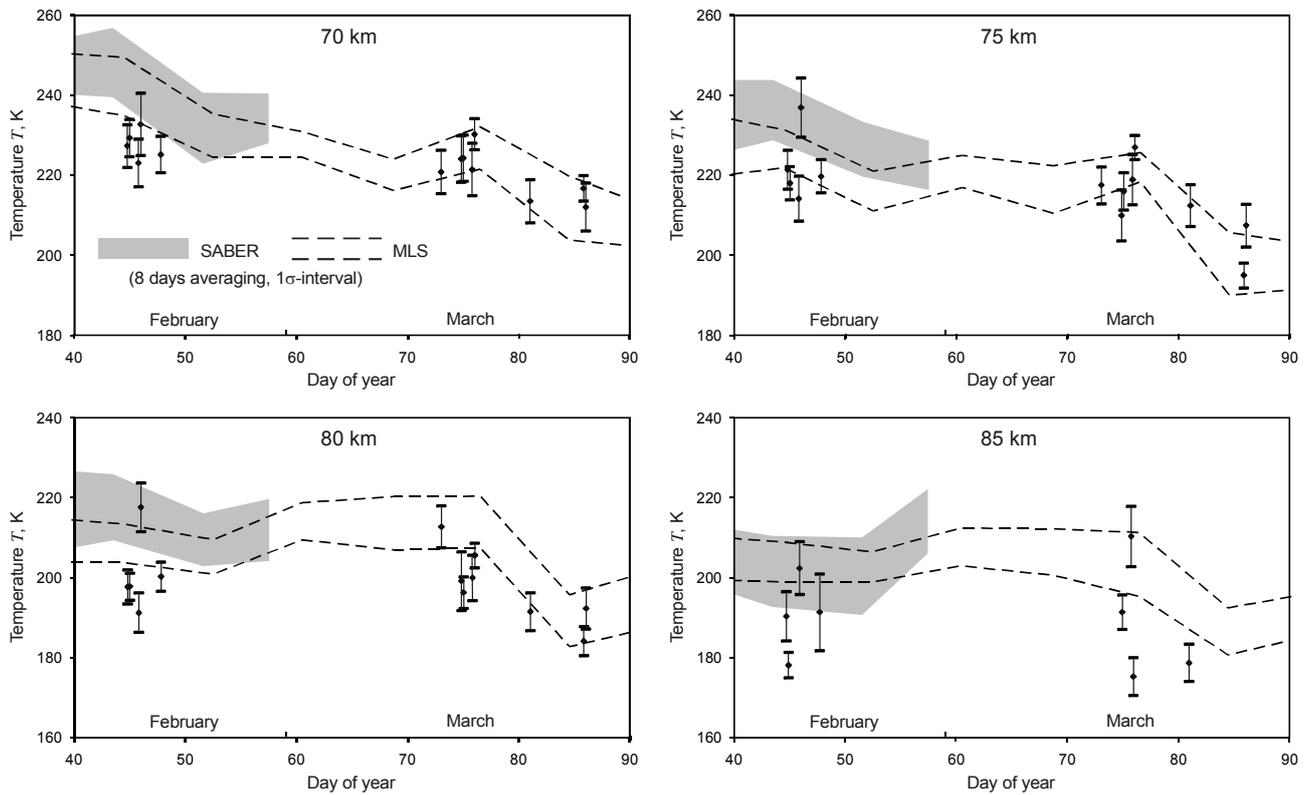

*Figure 3. Mesospheric temperature by twilight measurements compared with satellite data.*

Figure 3 shows the temperature values at the same altitudes for all the observed twilights in February and March, 2015 (the dots). Just as in [Ugolnikov and Maslov, 2013b], these values are compared to the TIMED/SABER and EOS Aura/MLS data, averaged by ±3° in latitude, ±10° in longitude and ±4 days in time (the 1σ-intervals are shown). Here, we again see a close agreement. Figure 4 presents the experimental temperature profiles averaged over the observations of mid-February, mid-March, and late March compared to the profiles provided by the same space experiments. The most remarkable feature of these profiles is a rapid drop of temperature at 70-80 km between the middle and late March, clearly observed in both the twilight and EOS Aura/MLS data.

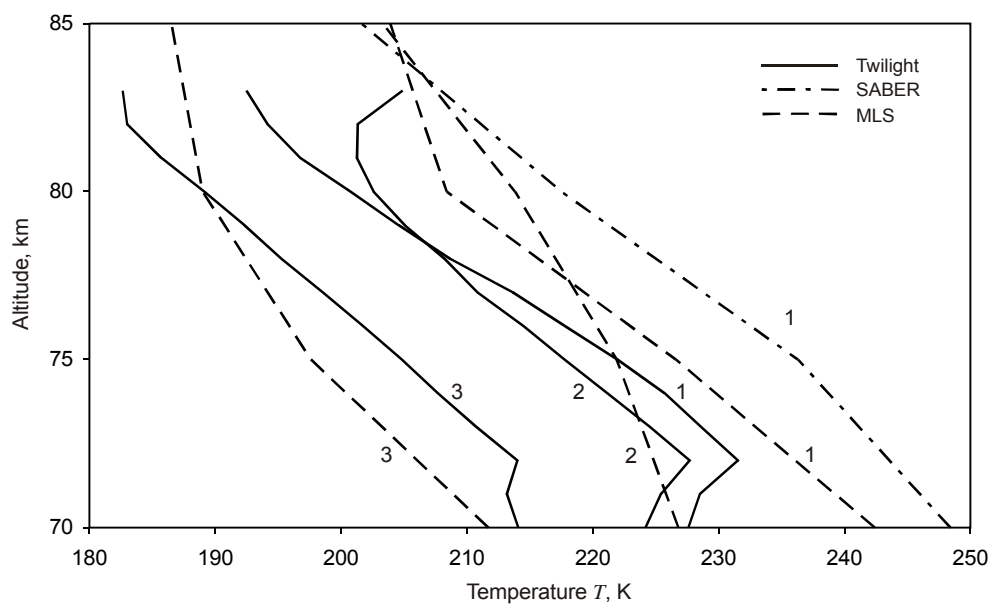

*Figure 4. Temperature profiles averaged by three observational periods compared with satellite data (1 - February, 13-16, 2 - March, 14-17, 3 - March, 22-27, 2015).*

We can also estimate the polarization of single scattering and compare it to the polarization of the Rayleigh scattering. This can be done in terms of the parameter

$$q_0 = q(\theta)/q_R(\theta) \qquad (2),$$

where $q$ and $q_R$ are the measured polarization and the polarization of the Rayleigh light scattering by the angle $\theta$ [Ugolnikov and Maslov, 2013b]. The values of $q_0$ averaged over the observable range of $\theta$ for the effective baseline altitude of the layer of 80 km were calculated by the same method and shown in Figure 5. These values are close to unity, indicating a low level of dust in the mesosphere. This is quite natural, since the northern polar mesosphere in late winter and spring is always located beyond the Earth relative to its orbital path, indicating a weak inflow of interplanetary dust. Large meteor showers are not active during this period of the year. This fact confirms the Rayleigh type of the single scattering in the mesosphere and validates the possibility to construct the temperature profiles based on these data.

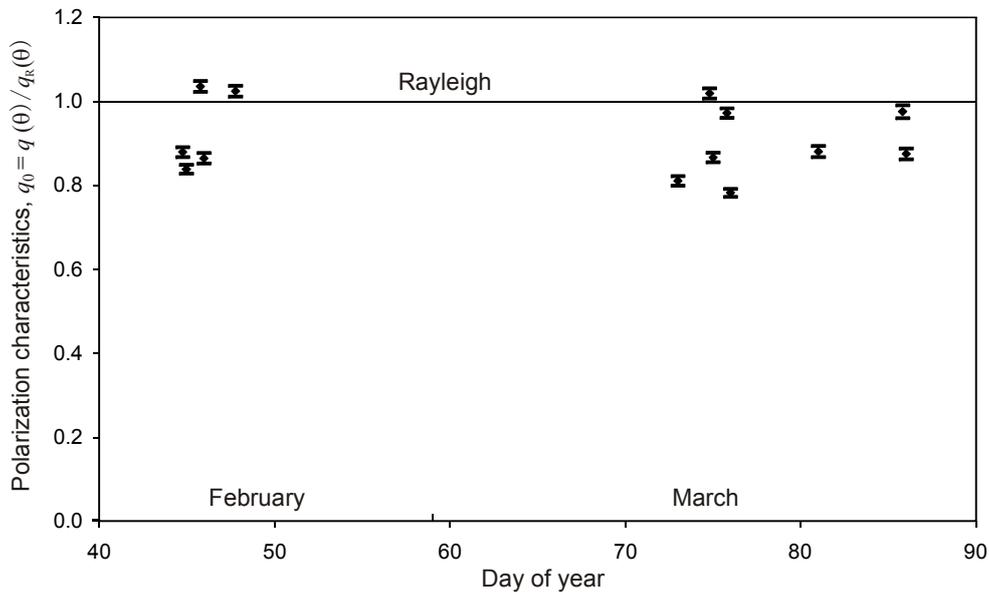

*Figure 5. Polarization characteristics of the single scattering above 80 km as the function of observation date.*

## 5. Discussion and conclusion

This paper presents the very first results of the mesosphere twilight sounding in a new observation site beyond the Polar Circle. Until recently, a similar study has only been successfully carried out at a single site close to Moscow. The observations in the cited experiment covered the late spring and summer seasons since 2011 [Ugolnikov and Maslov, 2013ab, 2014, 2015].

The high-latitude mesosphere during the spring transitional season, characterized by rapid changes of temperature, can be considered as the best correction test for the whole twilight technique. At the same time, this study is interesting since the polar mesosphere is likely to be most sensitive to the changes in the general atmospheric characteristics (including greenhouse gases). The long duration of the twilight makes it possible to increase the density of the data by the solar zenith angle, which is important since the baseline altitude of the effective layer rapidly increases with the increase in the depression of the Sun under the horizon. The same instruments can be used for the aurora observations (including the polarization measurements) during the night. However, the midnight sun and light nights during the summer cause a seasonal break in this work.

The analysis of the first data demonstrates the agreement between the temperature estimates with the satellite measurements and natural levels of single scattering polarization in the upper mesosphere. This validates the efficiency of twilight sounding for temperature and dust analysis during different seasons and at different locations. These observations can even be conducted in the urban conditions with high night sky background, as it has been done in the present study.

**Acknowledgments**
We are grateful to Yu. Ivanov (Head Astronomical Observatory, Kiev, Ukraine) for production of high-quality fish-eye lens suitable for polarization measurements. We also thank T. Sergienko (Swedish Institute of Space Physics, Kiruna) and N. Shakhvorostova (Lebedev Physical Institute, Russia) for the help during the data procession.